\documentclass[prd, aps, superscriptaddress, preprintnumbers, twocolumn, floatfix, nofootinbib]{revtex4}

\usepackage{amsfonts}
\usepackage{amsmath}
\usepackage{amssymb}
\usepackage{bm}
\usepackage{dcolumn}
\usepackage{graphicx}   
\usepackage[latin1]{inputenc}
\usepackage{latexsym}
\usepackage{rotating}
\usepackage{hyperref}
\usepackage{graphicx}
\usepackage{color}
\usepackage{xcolor}

\usepackage{amsfonts}
\usepackage{amsmath}
\usepackage{amssymb}
\usepackage{bm}
\usepackage{dcolumn}
\usepackage{graphicx}
\usepackage[latin1]{inputenc}
\usepackage{latexsym}
\usepackage{rotating}
\usepackage{hyperref}

\newcommand\be{\begin{equation}}
\newcommand\ba{\begin{eqnarray}}
\newcommand\ee{\end{equation}}
\newcommand\ea{\end{eqnarray}}

\begin{document}

\title {Massive black holes at high redshifts  from superconducting cosmic strings}

\author{Bryce Cyr}
\email{bryce.cyr@mail.mcgill.ca}
\affiliation{Physics Department, McGill University, Montreal, QC, H3A 2T8, Canada}

\author{Hao Jiao}
\email{hao.jiao@mail.mcgill.ca}
\affiliation{Physics Department, McGill University, Montreal, QC, H3A 2T8, Canada}

\author{Robert Brandenberger}
\email{rhb@physics.mcgill.ca}
\affiliation{Physics Department, McGill University, Montreal, QC, H3A 2T8, Canada}

\date{\today}

\begin{abstract}

\noindent The observation of quasars at high redshifts presents a mystery in the theory of black hole formation. In order to source such objects, one often relies on the presence of \textit{heavy seeds} ($M \approx 10^{4-6} \, M_{\odot}$) in place at early times. Unfortunately, the formation of these heavy seeds are difficult to realize within the standard astrophysical context. Here, we investigate whether superconducting cosmic string loops can source sufficiently strong overdensities in the early universe to address this mystery. We review a set of \textit{direct collapse} conditions under which a primordial gas cloud will undergo monolithic collapse into a massive black hole (forming with a mass of $M_{BH} \approx 10^5 \, M_{\odot}$ at $z \approx 300$ in our scenario), and systematically show how superconducting cosmic string loops can satisfy such conditions in regions of the $G\mu-I$ parameter space.

\end{abstract}

\pacs{98.80.Cq}
\maketitle

\section{Introduction} 
The origin of the active galactic nuclei (AGN) at redshifts higher
than $z = 7$ which are believed to harbour super-massive black holes with masses up to $10^9 \, M_{\odot}$ is a growing mystery for astrophysics and cosmology \cite{WangBH, Vol1, Vol2, Vol3}. In order to form such an AGN, a nonlinear overdensity needs to be present at early times. Unless a prolonged period of super-Eddington accretion is invoked, then according to the standard cosmological model ($\Lambda$CDM), the number density of such nonlinear seeds at high redshifts is insufficient \cite{jerome} to explain the observed abundance of high redshift AGNs.

One way to resolve this problem is to consider a cosmological model with large primordial non-Gaussianities in the required mass range. A natural candidate for such non-Gaussianities are cosmic string loops which inevitably arise in a large class of particle physics models beyond the Standard Model. A causality argument \cite{Kibble1, Kibble2} ensures that if Nature is described by a field theory which has cosmic string solutions, then a network of strings will form in the early universe and persist to the present time. In \cite{jerome} it was pointed out that cosmic string loops could provide the seeds for the observed high redshift super-massive black holes. More recently \cite{IMBH}, we studied the distribution of such seeds and pointed out that the distribution of seeds can be consistent with both the number density of super-massive and mass-gap black holes.

A key question not addressed in the previous works is the mechanism by which the nonlinear fluctuations induced by a string loop can collapse to form a black hole. This is the problem we address in this paper. We show that in the case of superconducting strings, all of the criteria for {\it Direct Collapse Black Hole Formation} can be satisfied in a range of cosmic string parameter values.

In the following section, we give a brief review of cosmic strings, discussing current constraints and highlighting their potential as seeds of supermassive black holes at high redshifts. In Section III we study accretion onto superconducting string loops. Section IV follows with a systematic study of the direct collapse criteria, and a proof-of-concept on how superconducting strings can efficiently meet these criteria. We conclude in Section V, discussing possible implications of these early black holes, as well as limitations of our study that we plan to address in the future.

\section{Cosmic String Overview}

Topological defects are objects that may form at the interface of cosmological phase transitions, provided that the vacuum manifold after such a transition is degenerate, and not simply connected. In three spatial dimensions, there are three classes of defects that are stable (preserved by a topological charge), domain walls from disconnected vacua, cosmic strings from a $1$-sphere vacuum manifold, and monopoles from a $2$-sphere vacuum manifold. Additionally, a vacuum manifold with a $3$-sphere topology gives rise to cosmic textures which can source interesting phenomenological signatures (for example, in the cosmic microwave background \cite{textureBC} \cite{textureCOLD}) even though they are not stable. In this work, we focus on an observational implication of cosmic strings.

Cosmic strings are nearly one dimensional objects that may form in the very early universe (see e.g. \cite{CSreviews1, CSreviews2, CSreviews3} for in-depth reviews). If the symmetry breaking patterns at high energies admit cosmic string solutions, the Kibble mechanism \cite{Kibble1, Kibble2} states that they must form, and will permeate the universe. Strings have a small but finite width, whose interiors consist of a condensate of gauge and scalar fields which were present in the previously unbroken phase. An attractive feature of regular cosmic string theories is that they possess only one free parameter, $\mu$, the string tension. If the symmetry breaking scale for cosmic string formation is $\eta$, the tension is simply related by $\mu \sim \eta^2$. In the literature, this is usually expressed as the dimensionless quantity $G\mu$ where $G$ is Newton's constant. Certain symmetry breaking patterns (for example \cite{witten, OstWitten, VilSCS}) can give rise to cosmic strings which carry a current ($I$) and are therefore referred to as superconducting. In this work, we will remain agnostic to the microphysics of the symmetry breaking, and so the parameter space is extended to include the current as a second free parameter. Phenomenological signatures of strings are often sensitive to larger values of $G\mu$, and thus also offer a way to probe particle physics from the top-down, complementary to accelerator experiments such as the Large Hadron Collider \cite{RHBtop}.

If cosmic strings form, they exist as both long strings (with curvature radius greater than the Hubble radius), and a distribution of smaller (sub-Hubble) string loops which form from the intersections and self-intersections of the long strings. It is understood that the distribution of long strings follows a scaling relation, but the nature of the loop distribution is more uncertain. Nambu-Goto simulations model the strings as truly one dimensional, and find that the loops also follow a scaling relation \cite{NGsim1, NGsim2, NGsim3, NGsim4, NGsim5, NGsim6, NGsim7, NGsim8}, whereas Abelian-Higgs field theory simulations find a negligible fraction of energy in the loops \cite{AHsim1, AHsim2, AHsim3}. The origin of this discrepancy is a long-standing problem with no clear answer thus far. Our work represents a proof-of-concept and therefore makes no assumption on the properties of the underlying loop distribution, aside from requiring that loops form and persist for at least a Hubble time.

Originally, cosmic strings were thought as an alternative mechanism to seed perturbations observed in the cosmic microwave background (CMB), requiring a string tension of $G\mu \sim 10^{-6}$ \cite{stringSeeds1, stringSeeds2, stringSeeds3, stringSeeds4}. However, cosmic strings don't give rise to acoustic oscillations, and so the now detailed studies of the angular power spectrum of the CMB rule out cosmic strings as the dominant source of structure formation, setting a robust bound of $G\mu \leq 10^{-7}$ \cite{cmbBound11, cmbBound12, cmbBound13} \cite{cmbBound21, cmbBound22, cmbBound23}. Long strings can also source line discontinuities in CMB temperature maps, which have the possibility of strengthening this bound by a couple orders of magnitude \cite{stringLine11, stringLine12} \cite{stringLine21, stringLine22, stringLine23, stringLine24, stringLine25, stringLine26, stringLine27}. In addition to these bounds, a distribution of cosmic string loops will produce a stochastic background of gravitational waves \cite{GWcutoff} which could be detected by current generation pulsar timing arrays, or next generation gravitational wave observatories such as LISA \cite{LISA}. Recently, the NANOGrav \cite{NANOgrav1}, PPTA \cite{NANOgrav2}, and EPTA \cite{NANOgrav3} collaborations have reported on the detection of a common-spectrum process which has been validated by the International Pulsar Timing Array group \cite{NANOgrav4}. If this signal has a cosmic string origin, it may be sourced by a loop distribution with string tension $G\mu \sim 10^{-10}$ \cite{StringGWs}. Alternatively, if this signal is not due to string loops, a stronger bound of $G\mu \leq 10^{-11}$ can be placed on the string tension assuming that the loop distribution follows a scaling relation (as is seen in Nambu-Goto simulations).

Besides sourcing a gravitational wave background, string loops have been considered in a wide variety of contexts. Work has been done on modelling the progenitors of fast radio bursts (FRBs), where annihilations of microstructure (cusps) on both regular \cite{FRBBC} and superconducting loops \cite{SCFRB1, SCFRB2, SCFRB3, SCFRB4, SCFRB5, SCFRB6, SCFRB7} could source the signal. In addition, these cusp annihilations may also influence the global $21$-cm signal by injecting additional photons into the background, heating it up. While this effect is small for ordinary string loops \cite{21cmReg}, new regions of parameter space were constrained in the context of superconducting loops \cite{21cmSCS}. String loops are regions of trapped energy, and thus are also capable of sourcing significant gravitational potentials at early times. The high redshift formation of galaxies and stars can lead to constraints coming from early reionization and structure formation \cite{ERolum} \cite{ERloeb} \cite{ERSCS}, though these are roughly the same as the CMB bound, $G\mu \leq 10^{-7}$. Accretion onto loops was studied in detail in \cite{UCMH}, where the formation of ultra-compact minihalos was considered.

Due to their gravitational potential, string loops have been proposed as the solution to some astrophysical mysteries, such as the origin of globular clusters \cite{lin1, lin2}, as well as supermassive and intermediate mass black holes \cite{jerome} \cite{IMBH}. Another astrophysical mystery is the observation of high redshift quasars, which are thought to host central black holes of mass $10^{8-9} \, M_{\odot}$  at redshifts $z \geq 7$ \cite{WangBH}. These objects are challenging to explain within the standard paradigm, as black holes formed through the typical stellar channels do not have enough time to accrete this much mass without a prolonged period of hyper-Eddington accretion \cite{Pacucci}. 

If string loops exist, they would be present at very high redshifts and persist through cosmic time, making them ideal candidates to source and seed these supermassive black holes (SMBHs). However, a question that has not been addressed clearly in recent proposals, is how the nonlinearity sourced by the string loop can collapse into a black hole at early times. In matching to observations, this is often assumed to be the case, and hence the main purpose of this work is to investigate if this collapse truly does take place. 

A cosmic string loop of radius $R$ has mass $M_{loop} = \beta \mu R$ (where $\beta = 2\pi$ refers to a circular loop). As string loops oscillate about their centre of mass, one can ask if a given loop ever satisfies the Hoop conjecture, that is, does its total mass ever reside within its Schwarzschild radius. For low string tensions ($G\mu \ll 10^{-3}$) the Hoop conjecture is satisfied for a  negligible fraction of string loops, leading to the conclusion that loop collapse into black holes is difficult to realize in nature \cite{sbh1, sbh2}. However, this need not be true when one considers not just the loop, but also its immediate environment. One should note that in full numerical relativity, it has been shown that for large values of the string tension ($G\mu \sim 10^{-2}$), it is possible for the loops to collapse on their own \cite{Lim1} \cite{Lim2}. However, such high tensions are ruled out for cosmologically stable strings. 

Much work has gone into attempting to source these high redshift SMBHs within a standard $\Lambda$CDM universe. One of the most promising avenues is the monolithic collapse of a pristine gas cloud, capable of producing an initially massive ($M_{BH,i} \approx 10^5 \, M_{\odot}$) black hole a mere million years after the collapse begins \cite{LoebDCBH1, LoebDCBH2, LoebDCBH3, LoebDCBH4, LoebDCBH5, LoebDCBH6} \cite{ZoltanDCBH1, ZoltanDCBH2, ZoltanDCBH3}. In order to produce a so-called \textit{Direct Collapse Black Hole} (DCBH), a number of favourable conditions within the gas cloud must be met. Unfortunately, one of the conditions (the presence of a strong Lyman-Werner background, see below for details) is exceedingly difficult to satisfy given only standard $\Lambda$CDM (see \cite{ZoltanDCBH1, ZoltanDCBH2, ZoltanDCBH3} and references therein). In this work, we show that for superconducting cosmic strings, there is a region of parameter space which satisfies all of the DCBH criteria and therefore may allow for the early formation of high mass black holes.

\section{Accretion onto Superconducting Cosmic String Loops}

String loops typically form with curvature radii some fraction of the Hubble radius, $R \sim \alpha t_f$, where $t_f$ is the time of loop formation and $\alpha \sim \mathcal{O}(0.1)$ (see \cite{CSreviews1, CSreviews2, CSreviews3} for details). After formation, these loops gradually lose energy through a variety of mechanisms which produce gravitational waves and particle content, such as cusp annihilations, kinks, and the overall oscillatory motion of the loop. For values of the string tension and current that we are most interested in, $G\mu \geq 10^{-17}$, the production of gravitational waves is the dominant energy loss mechanism for the loops \cite{GWcutoff} (hence why, for a given loop distribution, one gets strong constraints from the non-detection of a stochastic gravitational wave background from pulsar timing). As the strength of these processes depend on the radius of the loop, one can define a gravitational cutoff radius at time $t$ as $R^{GW}_c(t) = \gamma \beta^{-1} G\mu t$ (where $\gamma \sim \mathcal{O}(100)$ from simulations), where loops with $R < R^{GW}_c$ will lose all of their energy and decay within one Hubble time. We use the ``sudden decay" approximation, in which a loop with radius $R > R_{c}^{GW}$ maintains its size until it no longer satisfies this condition.

If the strings are superconducting (that is, the loops carry a current $I$), their direct coupling to electromagnetism enhances the energy loss to photons and charged fermions \cite{SCFRB4}. One can define a critical current, $I_c = \gamma \kappa^{-1} (G\mu)^{3/2} m_{pl}$ (where $m_{pl}$ is the Planck mass and $\kappa \sim \mathcal{O}(1)$), and if the current on the string is $I > I_c$, the electromagnetic decay channel is dominant. The value of the cutoff radius depends on this current through

\begin{align}
R_c(t) = \begin{cases}
R_c^{GW}(t) = \gamma \beta^{-1} G\mu t \hspace{19mm} \textrm{ $I < I_c$}\\
R_c^{EM}(t) =  \kappa \beta^{-1} \frac{I}{m_{pl}} (G\mu)^{-1/2} t \hspace{4mm} \textrm{ $I \geq I_c$}.
\end{cases}
\end{align}

\noindent String loops act as early non-linearities in the density field, and will accrete the surrounding dark matter and baryons. A detailed study of this accretion was performed in \cite{UCMH} for string loops forming and persisting during the radiation and matter dominated eras, with applications to ultra-compact minihalos in mind. 

We adopt their analysis, which describes the spherical infall of matter from the nearly smooth (linear) background density field at early times. As one would expect, string loops that exist in the radiation dominated era accrete at a highly inefficient rate, yielding a total non-linear mass at $t_{eq}$ of at most twice the loop mass, depending on if the loop had decayed by that point. However, once a loop enters the matter dominated epoch, its dark matter halo grows steadily, following the expression 

\begin{align} \label{Growth}
M_{DM}(z) = M_{loop} \left(\frac{z_{eq}+1}{z+1}\right) \hspace{8mm} (z \leq z_{eq}).
\end{align}

\noindent This growth occurs through the decoupling of spherical shells of dark matter from the Hubble flow, turning around at a physical distance from the centre of the loop, $r_{TA}$. These shells fall to the centre of the halo on free-fall timescales, during which many shell-crossings occur, and the matter subsequently virializes. The density profiles of these string-seeded halos follow a $\rho \sim r^{-9/4}$ profile \cite{CSprofile1, CSprofile2, CSprofile3}, steeper than the usual $\rho \sim r^{-2}$ observed from Gaussian fluctuations \cite{Gaussprofile}. Heuristically, the physical picture for the formation of an early massive gas cloud is this:

\begin{itemize}
\item String loops are formed before matter-radiation equality and accrete at a negligible rate.
\item At $t_{eq}$, dark matter begins falling into the potential wells sourced by the string loops. However, before recombination the Jeans mass is very large ($M_J \approx 10^{12} \, M_{\odot}$) and so the baryonic matter remains unbound to the newly forming halo.
\item After recombination the Jeans mass drops rapidly (to roughly $\approx 10^5 \, M_{\odot}$), and baryons collapse into string-seeded dark matter halos satisfying $M(z) > M_J$, where they virialize to a uniform temperature, $T_{vir}$. We show a more detailed computation of this in the following section.
\end{itemize}

These accretion dynamics were originally formulated for a non-superconducting cosmic string. In Appendix A, we review the basic arguments of \cite{UCMH} more quantitatively, and show that the additional radiation pressure produced by a superconducting cosmic string doesn't perturb the picture described above in any major way. We will now show that a gas cloud formed in this way is capable of satisfying conditions which would lead to its monolithic collapse into a massive, high redshift black hole.

\section{Direct Collapse Conditions}

The formation of stellar mass black holes in our universe is a fairly well understood process: once the internal temperature of a star is no longer high enough to mediate nuclear interactions, the star collapses under its own gravitational force, often leading to a supernova and a remnant. For sufficiently massive stars, this remnant cannot be sustained by neutron degeneracy pressure and collapses into a black hole \cite{stellarForm}. The LIGO-Virgo collaboration has confirmed that these stellar mass ($\mathcal{O}(1) \, M_{\odot} \, \leq \, M_{BH} \, \leq \, \mathcal{O}(100)\, M_{\odot}$) black holes are indeed ubiquitous in nature \cite{LIGO1, LIGO2, LIGO3}. 

However, observations of high redshift quasars suggest that supermassive black holes are already in place at early times. The most robust high redshift black hole observation to date has a mass of $1.6 \times 10^9 \, M_{\odot}$ at $z = 7.642$ \cite{WangBH}, although there have recently been observations of two galaxy candidates \cite{HarikaneBH} at $z \sim 13$, which may be interpreted as quasars hosting black holes of order $M_{BH} \approx 10^8 \, M_{\odot}$ \cite{PratLoeb}. Conservatively, a population of ``light" black hole seeds (initial mass $M_{BH} \sim 10^2 \, M_{\odot}$) will form through the collapse of Pop III stars. However, there is simply not enough time for them to grow to these supermassive objects by $z \sim \mathcal{O}(10)$ unless super or hyper-Eddington accretion can be reliably sustained (see \cite{Pacucci} and references therein).

The formation of ``heavy" seeds (initial mass $M_{BH} \sim 10^{4-6} \, M_{\odot}$) on the other hand, offers a more realistic way to form supermassive black holes at high redshift. These heavy seeds can be formed through runaway merger processes of lighter black holes or Pop III stars in highly dense environments \cite{PopIIImerge1, PopIIImerge2}, or through the monolithic collapse of a metal-free gas cloud \cite{LoebDCBH1, LoebDCBH2, LoebDCBH3, LoebDCBH4, LoebDCBH5, LoebDCBH6} \cite{ZoltanDCBH1, ZoltanDCBH2, ZoltanDCBH3}. Unfortunately, both of these formation channels are still quite difficult to realize in the standard cosmological and astrophysical context, and hence the origin of high redshift SMBHs still remains an open question.

In order for a gas cloud to collapse without fragmentation, a number of conditions must be met,

\begin{itemize}
\item The cloud must contain sufficient material to collapse and feed an initial protostar or seed black hole. If we want the final seed mass to be $M\approx 10^5 \, M_{\odot}$, we require at least this much baryonic matter in our string-seeded overdensities.
\item The atomic cooling threshold of the halo must be met in order to trigger the collapse without fragmentation. This requires the virial temperature of the halo to be $T_{vir} = 10^4 \, \textrm{K}$ (or higher).
\item In order to reach the atomic cooling threshold, low temperature cooling pathways must be suppressed. Low temperature cooling occurs via heavy elements or molecular hydrogen being present in the halo. Therefore, we demand that our early formed halos haven't undergone star formation that would pollute it with metals. Additionally, a Lyman-Werner (LW) background of $J_c \approx 10^{-44} - 10^{-42} \, \textrm{GeV}^3$ is required to quench the formation of molecular hydrogen.
\end{itemize}

If these conditions are met, a direct collapse event may occur. In such an event, the gas cloud collapses nearly isothermally, forming a rapidly accreting protostar at the centre\footnote{see figures 7, 9 and 11 of \cite{ZoltanDCBH1, ZoltanDCBH2, ZoltanDCBH3} as well as \cite{oneZone} \cite{numCollapse1} \cite{numCollapse2} for detailed simulations and semi-numerical modelling of this process.}. Following the work of \cite{ZoltanDCBH1, ZoltanDCBH2, ZoltanDCBH3}, this central protostar is initially quite light ($M_i \approx 0.2 \, M_{\odot}$), but  accretes at a rate of roughly $\dot{M} \approx 20 c_s^3/G \approx 1 \, M_{\odot}/\textrm{yr}$ \cite{numCollapse2} provided that the temperature remains as high as $T_{vir} \sim 10^4 \, \textrm{K}$ during the collapse. At this accretion rate, the protostar exhibits an instability at roughly $M \approx 2 - 3.5 \times 10^5 \, M_{\odot}$, collapsing into a black hole of similar mass \cite{Umeda} \cite{Woods}.

It is clear that for standard astrophysical galaxies, the requirement for $H_2$ dissociation processes such as a strong Lyman-Werner background is difficult to realize. In principle, this background can be produced by star formation in the galaxy, though this would quickly pollute the galaxy with various metals, in violation with the \textit{pristine mass} condition. One possibility would be to source the necessary LW background from a nearby star-forming galaxy, though this also requires a tuned scenario.

If a superconducting cosmic string resides at the centre of these protogalaxies, however, there is the possibility of sourcing this background without requiring any star formation. Therefore, within a given region of the $G\mu-I$ parameter space, loops with a radius $R > R^{BH}_c$ (where $R^{BH}_c$ is the critical loop radius for black hole formation) will undergo a gravitational collapse. We emphasize that it is not the loop itself that is collapsing, but instead the baryonic matter in the string-seeded overdensity that forms the black hole. In the following subsections, we show that superconducting cosmic strings present ideal environments to source direct collapse black holes at very early times.

\subsection{Sufficient Pristine Mass}

In order to grow a protostar to the point that it can undergo a collapse into a black hole with mass $M_{BH} \approx 10^5 \, M_{\odot}$, we require at least that much pristine, baryonic material. As we are envisioning a picture in which these gas clouds collapse shortly after recombination, no star formation is expected to have occurred and so the requirement on pristine material is trivially satisfied. 

Eq. (\ref{Growth}) represents the growth rate of the dark matter halo after matter radiation equality. Baryons remain unbound to the halo until recombination due to a high Jeans mass. After recombination, however, the Jeans mass drops rapidly to a roughly constant value \cite{firstlight},

\begin{align}
M_j \approx 1.35 \times 10^5 \left(\frac{\Omega_m h^2}{0.15}\right)^{-1/2} \, M_{\odot}.
\end{align}

\noindent Dark matter halos with $M \geq M_J$ will begin capturing baryons quickly following recombination. The physical radius that a shell of matter which is beginning to collapse at a redshift $z$ would have in the absence of a string loop is given by the non-linear radius, $r_{nl}$,

\begin{align} \label{rnl}
r_{nl}(z) = \left( \frac{3 \beta \mu R}{4 \pi \rho(z)} \cdot \frac{1+z_{eq}}{1+z}\right)^{1/3}.
\end{align}

\noindent This is related to the turn-around radius by $r_{nl} = 2 r_{TA}$. Since we are at very early times, we assume that both the dark matter and baryonic density fields are smooth, and thus the total baryonic mass inside this radius is

\begin{align}
M_b(z) = \frac{\Omega_b(z)}{\Omega_{M}(z)} \beta \mu R \left( \frac{1+z_{eq}}{1+z}\right) \hspace{10mm} (z< z_{rec}).
\end{align}

\noindent The baryon fraction $\Omega_b/\Omega_{M} \approx 1/6$ and is roughly constant in time. If we require $M_b \geq 10^5 \, M_{\odot}$ in order to form direct collapse black holes, we can see that the condition for baryonic infall ($M_{DM}(z) \geq M_J(z)$) is automatically satisfied. Therefore, in order to have enough pristine baryonic matter in the halo, we require

\begin{align}
G\mu \geq \frac{\Omega_{M}}{\Omega_b} \frac{G}{\beta} \frac{1}{R} \left( \frac{z+1}{z_{eq}+1}\right) \times 10^5 \, M_{\odot}
\end{align}

\noindent or, in units of $c = \hbar = k_b = 1$ with $\beta = 10$

\begin{align} \label{massCon}
G\mu \geq  2 \times 10^{-13} \left( \frac{t_{eq}}{R} \right) \left(\frac{z+1}{z_{eq}+1}\right).
\end{align}

\noindent Recall that at any time $t$ (or redshift $z$) loops form with radius $R_{form} = \alpha t$, and loops with gravitational cutoff radius $R_c^{GW} = \gamma \beta^{-1} G\mu t$ decay away (we will be concerned with values of the current $I<I_c$, meaning that gravitational emission is the dominant decay mechanism). Expressing $t$ in terms of $t_{eq}$ this means that during the matter epoch loops exist with a range of radii

\begin{align}
\gamma \beta^{-1} G\mu \left( \frac{1+z_{eq}}{1+z}\right)^{3/2} t_{eq} \, \leq \, R \, \leq \, \alpha \left( \frac{1+z_{eq}}{1+z}\right)^{3/2} t_{eq}.
\end{align}

\noindent However, we need our loops to be in place by $t_{eq}$ so that they may grow their dark matter halos until recombination, therefore our range of accessible string radii is

\begin{align}
\gamma \beta^{-1} G\mu t_{eq} \, \leq \, R \, \leq \, \alpha t_{eq}.
\end{align}

\noindent We will be most interested in the case where the baryons fall into the halo right after recombination. Therefore, some fraction of the loop distribution with $R > R^{BH}_c$ will attract $M_b \geq 10^5 \, M_{\odot}$ in the initial baryonic collapse at $z_{rec}$. The value of $R^{BH}_c$ can be computed by rearranging (\ref{massCon}) for a given value of $G\mu$

\begin{align} \label{critLoop}
R^{BH}_c = \frac{2 \times 10^{-13}}{G\mu} t_{eq} \left(\frac{z+1}{z_{eq}+1}\right).
\end{align}

\noindent At a given redshift, this implies that loops with radii in the range

\begin{align}
R_C^{BH} \leq R \leq \alpha t_{eq}
\end{align}

\noindent will accrete a sufficient amount of gas to collapse into a $10^5 \, M_{\odot}$ black hole.

Note that the fact that the strings are superconducting is irrelevant for this condition, so both ordinary and superconducting cosmic strings are capable of sourcing massive gas clouds in the early universe (as was considered in eg. \cite{ERolum} \cite{ERloeb}).

\subsection{Atomic Cooling Threshold}

The second criterion required to kickstart a direct collapse procedure is the formation of an atomic cooling halo. An atomic cooling halo is characterized by its inability to cool at low temperatures ($T_{vir} \leq 10^4 \, \textrm{K}$) through the usual pathways of molecular hydrogen or other heavy elements. As a result, the halo is unable to cool until it reaches a gas temperature of $T_{vir} \approx 8000 \, \textrm{K}$  when cooling via atomic line transitions in hydrogen begins. The lack of heavy elements in string-seeded overdensities was justified in the previous section, and we will discuss the quenching of $H_2$ in the following section. Here we compute the virial temperature of a string-seeded halo and show that it is above the atomic cooling threshold, implying that a collapse will indeed take place, provided that the halo satisfies the \textit{sufficient mass} criteria above.

As mentioned above, after recombination the Jeans mass drops rapidly and baryons fall into the halo. Again, we consider the spherical collapse model where shells of baryonic and dark matter turn around and fall towards the centre. The typical timescale for this infall is the free-fall time. When the mass of the infalling shell is much less than the halo mass, it can be approximated by

\begin{align}
t_{ff} \approx \sqrt{\frac{3\pi}{32 G \bar{\rho}}},
\end{align}

\noindent where $\bar{\rho}$ is the background energy density. Using the critical density, one can find that the free-fall time is roughly a Hubble time, $t_{ff} \approx (\pi/2) H^{-1}$. It is important to note that during a free-fall time, the collapsing shells of matter will cross one another, interacting and bringing the halo into an equilibrium \cite{shellCrossing}. Therefore, $t_{ff}$ is identified with the virialization timescale, when the gas in the halo reaches a uniform temperature, $T_{vir}$ which we will now compute. For clarity, a shell turning around at recombination has a free-fall time of $t_{ff}(z_{rec}) \approx 0.85 \, \textrm{Myr}$, implying that it forms a virialized halo at roughly $z \approx 500$.

The virial theorem provides a relationship between the gravitational and thermal energy of a halo

\begin{align}
\frac{G m_h M_{nl}}{r_{vir}} = \frac{3}{2} T_{vir},
\end{align}

\noindent where $m_h$ is the mass of hydrogen (the dominant baryonic constituent), $M_{nl}$ is the total nonlinear mass in a sphere of radius $r_{nl}$, and $r_{vir}$ is the virial radius of the halo, related to the turn-around and non-linear radii by $r_{vir} = (1/2) r_{TA} = (1/4) r_{nl}$. Recall that eq (\ref{rnl}) provides us with the redshift at which a certain shell becomes nonlinear, allowing us to determine the virial temperature of a string-seeded halo simply as a function of redshift. After evaluating numerical factors and setting $\beta = 10$, we find the virial temperature of a halo at redshift $z$ to be

\begin{align}
T_{vir}(z) \approx 8 \times 10^{13} \, \textrm{K} \,(G\mu)^{2/3} \left( \frac{R}{t_{eq}}\right)^{2/3} \left( \frac{1+z}{1+z_{eq}}\right)^{1/3}.
\end{align}

\noindent It is important to note that this expression must be evaluated at the redshift when a particular shell goes nonlinear. Therefore, even though shells that go nonlinear at recombination don't virialize until $z \approx 500$, their temperature must be evaluated at $z_{rec}$. The reason being that shells decoupling from the Hubble flow after recombination have not undergone sufficient shell crossings at $z = 500$ to virialize and thus don't contribute to the temperature.

In order to trigger atomic cooling, our halo must reach $T_{vir} \geq 8000 \, \textrm{K}$ allowing us to once again relate the string tension with the radius of a particular loop. This yields the constraint

\begin{align}
G\mu \geq 9.7 \times 10^{-16} \left( \frac{t_{eq}}{R}\right) \left(\frac{1+z_{eq}}{1+z}\right)^{1/2}.
\end{align}

\noindent This constraint is degenerate with the sufficient mass conditions. As the halo grows in mass, its virial temperature drops modestly as energy is diffused from the hot inner regions of the halo to the colder shells which fall in later. At a redshift of $z \sim 90$ this constraint becomes the dominant one. We are primarily concerned with the baryonic matter falling in shortly after recombination, and as a result, if the mass constraint is satisfied, the virial temperature of the halo will automatically be above the atomic cooling threshold. The critical loop radius will remain the same as in eq (\ref{critLoop}), and this condition is also not sensitive to the superconducting nature of our string loops.

\subsection{Lyman-Werner Background}
If a gas cloud becomes massive enough, structure formation theory tells us that it will collapse and fragment, leading to abundant star formation. In order for this fragmentation to occur, the gas must remain relatively cool ($T \leq 10^4 \, \textrm{K}$). Low temperature cooling in gas clouds can occur when heavy elements are present, which are usually sourced by an epoch of prior star formation in the halo. In a string-seeded halo, no such star formation has taken place, and so our gas is pristine, consisting of only hydrogen atoms.

However, cooling can also occur via the collisional excitation of molecular hydrogen, unless a sufficiently strong flux of Lyman-Werner photons ($E_{\gamma} \approx 10-13 \, \textrm{eV}$) is present. Lyman-Werner photons possess enough energy to quickly dissociate any newly formed $H_2$ allowing the halo to heat up to the atomic cooling threshold ($T \approx 10^4 \, \textrm{K}$), where the fragmentation of a collapsing gas cloud no longer occurs. The typical flux necessary to properly irradiate an optically thick, collapsing gas cloud is roughly \cite{ZoltanDCBH1, ZoltanDCBH2, ZoltanDCBH3}

\begin{align}
J_c &\approx (10^{-18} - 10^{-16}) \, \textrm{erg s}^{-1} \textrm{cm}^{-2} \textrm{Hz}^{-1} \, \textrm{Sr}^{-1} \nonumber \\
&\approx 2 \times ( 10^{-44} - 10^{-42}) \, \textrm{GeV}^3,
\end{align}

\noindent where we quote the second line in natural units. 

This condition, which is necessary for the monolithic collapse of a gas cloud, is notoriously difficult to realize within standard astrophysics and cosmology. Lyman-Werner backgrounds are usually sourced by the first stars in a halo. This presents a problem, as it also introduces metals into the halo, and therefore will not prevent cooling. Superconducting cosmic strings, on the other hand, can source such a background without star formation.

Superconducting strings carry a current, $I$, and are capable of emitting large bursts of electromagnetic radiation. The energy injection spectrum of a loop with radius $R$ and current $I$ is given by \cite{SCFRB4}

\begin{align} \label{SCSpow}
\frac{dP}{d\omega} = \kappa I^2 R^{1/3} \omega^{-2/3}.
\end{align}

\noindent To estimate the flux in Lyman-Werner photons, we consider an optically thick gas which traps photons emitted by the string (in the band $10 \, \textrm{eV}\, \leq \, E_{\gamma} \, \leq \, 13 \, \textrm{eV}$) at times after recombination. By integrating (\ref{SCSpow}) over a Hubble time at redshift $z$ and dividing by the virialized volume we obtain the following expression for the energy density per unit frequency at the frequency $\omega$

\begin{align}
\frac{d\rho}{d\omega} &\approx \frac{32}{3\pi} \frac{\kappa}{\beta} I^2 (G\mu)^{-2} R^{-2/3} \omega^{-2/3} t_{eq}^{-1} \nonumber \\
&\times \left(\frac{1+z}{1+z_{eq}}\right)^{5/2} \left( 1- \left(\frac{1+z}{1+z_{rec}}\right)^{3/2}\right) .
\end{align}

\noindent Expressing the current in units of the critical current $I_c$, and inserting values for the Lyman-Werner frequency $\omega$, the Planck mass and $t_{eq}$ we obtain flux of

\begin{align}
J \approx 10^{-34} &\left(\frac{G\mu }{10^{-10}}\right)^2\left(\frac{I}{I_c}\right)^2 \left(\frac{R}{t_{eq}}\right)^{-2/3} \left(\frac{1+z}{1+z_{eq}}\right)^{5/2} \nonumber\\
&\times \left( 1-\left(\frac{1+z}{1+z_{rec}}\right)^{3/2} \right)  \, \textrm{GeV}^3,
\end{align}

\noindent where we have evaluated the numerical constants as $\kappa = 1$ and $\gamma = 100$.  Shells of matter turning around at recombination will virialize and begin to collapse at $z \approx 500$, and so we evaluate at this redshift. Recall that for $I < I_c$ the dominant energy loss mechanism for the strings is gravitational waves, while for $I > I_c$ it is electromagnetic emission. By requiring that we have a flux $J \geq J_c \approx 10^{-43} \, \textrm{GeV}^3$, we find a constraint in the $G\mu$-$I$ plane

\begin{align} \label{SCSLW}
G\mu \geq 4 \times 10^{-14} \left( \frac{I}{I_c}\right)^{-1} \left(\frac{R}{t_{eq}}\right)^{1/3},
\end{align}

\noindent when we evaluate at $z = 500$, the black hole collapse redshift. Recall that any string loop with $R_c^{BH} \leq R \leq \alpha t_{eq}$ is capable of accreting enough mass to satisfy the direct collapse criteria. One may rightly worry that the radiation from a superconducting string could induce a strong pressure force which disturbs the accretion dynamics. We refer the reader to Appendix A where we consider these effects in more detail. By requiring that these pressure forces remain small, a complementary upper bound can be set on the string tension

\begin{align} \label{SCSRP}
G\mu \leq 10^{-2} \left( \frac{I}{I_c}\right)^{-3}  \left( \frac{R}{t_{eq}}\right)^5.
\end{align}

\noindent For a given loop radius $R$ which satisfies the sufficient mass constraint, there exists a wedge in the $G\mu$-$I$ plane which can produce a strong enough Lyman-Werner background, while not sourcing enough radiation pressure to disturb the accretion. 

Thus far we have remained agnostic to whether these loops exist as part of a scaling loop distribution as indicated by Nambu-Goto simulations, or whether they exist in relative isolation with no scaling solution and a negligible cosmological abundance. However, we must now make a comment, as if string loops are abundant, many authors (eg: \cite{ERSCS, 21cmSCS, SCSconstraints}) have attempted to constrain the $G\mu$-$I$ parameter space. In the case where loops are not abundant, these constraints generally don't apply and one may view the wedge formed by (\ref{SCSLW}) and (\ref{SCSRP}) as the allowed region of parameter space. 

If loops are abundant, it would appear from Fig. 8 of \cite{SCSconstraints} that this current wedge is already constrained. However, there exist a number of caveats that make it difficult to do a straightforward comparison between our work and the usual constraints in the literature. 

First, these constraints assume that all loops in the network are superconducting, with the same uniform current $I$, regardless of loop size. Early work on superconducting cosmic strings pointed out that the current generated on a string is dependent on the local external magnetic field at the time and place of loop formation \cite{witten, OstWitten, VilSCS}. Therefore, as a proof-of-concept work, it is possible that significant fluctuations in the magnetic field on length scales comparable to the radius of the loop could induce anomalously large currents on a small fraction of the loop distribution. This would source a strong Lyman-Werner background in this small fraction of loops, satisfying the criterion for direct collapse black hole formation, while remaining consistent with the constraints from \cite{SCSconstraints}.

Furthermore, if one has a scaling solution of loops, the observational signatures are often dominated by loops with the smallest radii, as they are the most abundant. The CMB anisotropy constraints \cite{SCSconstraints} overlap significantly with our target parameter values, but these constraints are strongly dominated by string loops that decay at recombination. The loops we consider here are necessarily much larger, as we require them not to decay before redshifts of at least $z \approx 500$ (in the case of mass shells turning around at recombination). This implies that the currents on our larger loops need not be as small as those on the ones decaying at recombination. In fact, one might also expect that since these two populations of loops form at different times, they will likely possess different currents as the external magnetic fields evolve between their formation times.

Finally, there is also the possibility that if a scaling loop distribution exists, not all of the loops need to be superconducting. Therefore, the strength of the constraints in the $G\mu$-$I$ plane will scale depending on the fraction of superconducting loops in the network, and the characteristic radii of the loops that do carry a current. Note also that we consider only the Lyman-Werner photons from the central loop. If a distribution of loops exists, a number of smaller loops will also be present in each halo and will also contribute to the dissociating photon background.

We conclude this section by remarking that regardless of whether a string loop belongs to an abundant distribution, or is produced in relative isolation, equations (\ref{SCSLW}) and (\ref{SCSRP}) provide upper and lower bounds for the string tension given a current and loop radius. This maps out a wedge in parameter space where a strong enough Lyman-Werner background is present to dissociate molecular hydrogen, but the radiation pressure does not significantly perturb the accretion dynamics.

\section{Discussion and Conclusions}

The progenitors of black holes at high redshifts inferred through quasar observations remains a tantalizing mystery in astrophysics and cosmology. At the moment, the most intuitive candidate is to invoke hyper-Eddington growth onto a stellar mass black hole formed through the death of a Pop III star. However, the conditions necessary for such rapid accretion are not well understood, and strong feedback will likely cause severe limitations on the growth rates of such objects. 

Another hypothesis is to invoke the direct collapse of a massive gas cloud at earlier times, as then only slow growth is necessary to grow them from an initial mass of roughly $M_{i} \approx 10^5 \, M_{\odot}$ to their observed masses of $M_{f} \approx 10^{8-9} \, M_{\odot}$ at redshifts of $z\approx 8-13$. The conditions for such a collapse have been known for some time: a gas cloud with sufficient mass, and a lack of cooling pathways at low temperatures will collapse without fragmentation when it reaches the atomic cooling threshold. Unfortunately, it is difficult for typical gaseous halos to satisfy these conditions simultaneously. However, in the presence of a superconducting cosmic string loop, we have shown that these conditions can indeed be satisfied.

What we have presented above is a proof-of-concept work showing that if superconducting cosmic strings exist, they act as nonlinear seeds at early times (around matter-radiation equality). For a range of parameters, they first grow a dark matter halo at $t_{eq}$, and begin accreting gas shortly after recombination. Using the spherical collapse model, we have argued that a virialized halo is in place roughly one Hubble time after the turnaround time for the first shell of matter (namely, at around $z \approx 500$). Electromagnetic emission from the superconducting string sources a strong Lyman-Werner background, ensuring that no molecular hydrogen is present. Therefore, these string-seeded halos can efficiently satisfy all of the direct collapse criteria, and may collapse into black holes. The time between the initial monolithic collapse of the gas cloud, and the $10^5 \, M_{\odot}$ black hole formation is about $1 \, \textrm{Myr}$, implying that these black holes have formed by $z \approx 300$. 

We have decided not to speculate on whether these string seeded black holes can match the observed abundance of high redshift quasars. We leave this to future work due to a number of caveats and subtleties that arise when applying this idea to a network of string loops. As mentioned in the previous subsection, the loops that can give rise to these halos are necessarily quite large ($R \gg R_c^{GW}$). Constraints on the $G\mu - I$ plane (see \cite{SCSconstraints}) are usually sourced mainly by the smallest loops, since they are the most abundant. As a result, one needs to revisit how sensitive these constraints (particularly the ones coming from CMB anisotropies) are to currents on the larger loops before directly applying them to our model. Also, as currents on the loops are generated at the time of formation by local electromagnetic fields, it is likely that loops with different radii have different currents. Large fluctuations of the magnetic field on very small scales may also cause some loops to develop an unusually strong current. Each of these nuances must be understood before a proper abundance computation can be performed for these string-seeded black holes.

Furthermore, we have assumed spherical accretion in the rest frame of the string loop. While this will likely be a good approximation for some fraction of a string loop network, in general one should consider the loop velocity distribution. The relative velocity between the loop and its surrounding medium weakens the efficiency of accretion, making it more difficult to fulfil the \textit{sufficient mass} criterion above. Historically, stringent bounds on the cosmic string tension $G\mu$ have been made weaker by considering this effect, such as in the case of early star formation \cite{ERloeb} and ultra-compact minihalos \cite{UCMH} from string loops. 

We have presented a proof-of-concept to show that in the regions of parameter space outlined by equations (\ref{massCon}), (\ref{SCSLW}), and (\ref{SCSRP}), superconducting string loops can seed gaseous halos which will collapse into early time supermassive black holes. While this represents a possible resolution of the high redshift quasar mystery, one should be cautious in applying these results to a network of string loops without addressing our above concerns. Nevertheless, this represents an important step in validating phenomenological models of black hole formation from cosmic strings.\\

\section*{Acknowledgements}
The research at McGill is supported in part by funds from NSERC and from the Canada Research Chair program. BC is grateful for support from a Vanier-CGS doctoral scholarship. HJ is supported in part by a Trottier award from the McGill Space institute. We would like to thank Pratika Dayal and Maxime Trebitsch for valuable discussions at an early stage, as well as Marta Volonteri and Patrick Peter for correspondence and encouragement.

\section*{Appendix A - Accretion Dynamics and Pressure Forces}

Here we summarize the main points of ref. \cite{UCMH} which explored the detailed accretion dynamics of dark matter onto cosmic string loops. Both our work and theirs adopt the ``sudden decay" approximation, in which the slow gravitational decay of the loop is neglected until the loop radius falls below the gravitational cutoff. This implies that once formed, loops keep their radius at $R_f = \alpha t_f$ until decay. The time of decay is then related to the time of formation by 

\begin{align}
t_d = \frac{\alpha}{\gamma G\mu} t_f.
\end{align}

\noindent The authors of \cite{UCMH} then consider the spherical collapse \cite{zeld} of dark matter onto initially isothermal fluctuations sourced by the string loops. Spherical shells are considered a physical distance $r(x)$ from the centre of the loop with 

\begin{align}
r(x) = a(x) b(x) \zeta_i, \hspace{9mm} x=  \frac{a(t)}{a(t_{eq})},
\end{align}

\noindent where $a(x)$ is the scale factor, $\zeta_i$ is the initial comoving coordinate associated with the shell at time $t_i$, and $b(x)$ compares this physical distance in the case with a loop present, and without. A shell is said to turn around when $\dot{r} = 0$, and so the rest of the work is dedicated to solving the dynamical equation for $b(x)$ in different regimes \cite{zeldKolb},

\begin{align}
x(x+1)\frac{d^2b}{dx^2}+ \left(1+\frac{3}{2}x\right)\frac{db}{dx} +\frac{1}{2} \left(\frac{1+\Phi}{b^2} - b \right) = 0.
\end{align}

\noindent The gravitational perturbation is simply $\Phi = \delta M/ M(r)$ where $\delta M = M_{loop}$ is the loop mass, and $M(r)$ is the total mass enclosed within physical radius $r$. 

For loops that form in the radiation era, it was found that even before $t_{eq}$, mass shells will turn around and fall onto the seed. If the loop forms and decays before $t_{eq}$, the total growth of the overdensity in the radiation era (with $x > x_d$) is given by

\begin{align}
M(x) = 2 x_d M_{loop} \frac{3x+2}{3x_d + 2} \hspace{9mm} ( x_d \, < \, x \, < \, 1),
\end{align}

\noindent where $x_d = a(t_d)/a(t_{eq})$ is the time when the loop decays. If the loop  has not decayed yet, the growth was found to be linear in $x$ at all times, following

\begin{align}
M(x) = 2 x M_{loop} \hspace{8mm} (x_f < 1 \,\, \textrm{and} \,\, 1<x<x_d).
\end{align}

\noindent Since at $t_{eq}$ we have $x=1$, this implies that a loop will build up an overdensity with mass $M(t_{eq}) = 2M_{loop}$ at matter-radiation equality. In our work we conservatively set the linear growth rate of the dark matter halo to be $M(x) = M_{loop} x$ in agreement with these results.

Superconducting strings produce a radiation field which can source an additional pressure force on the infalling matter. To estimate this effect, we consider the ratio of the gravitational binding force of a halo at redshift $z$ to the radiation pressure force coming from photons in that halo. For a density profile of the form $\rho(r) = \rho(r_{vir}) (r_{vir}/r)^{\alpha}$, the gravitational binding energy is 

\begin{align}
U_{grav} = - \frac{16 \pi^2 G}{(3-\alpha)(5-2\alpha)} \rho^2(r) r^5.
\end{align}

\noindent Under the conservative assumption that all electromagnetic radiation produced by the string loop between recombination and the redshift of matter infall (denoted here by $z$ where the physical distance from the centre of the loop to the shell is $r(z)$) remains bound to the halo, the radiation pressure is 

\begin{align}
P = \frac{3}{4\pi}\kappa I^2 r^{-3} (R \omega_c)^{1/3} &t_{eq} \left(\frac{1+z_{eq}}{1+z}\right)^{3/2} \nonumber\\
&\times \left( 1 - \left( \frac{1+z}{1+z_{rec}}\right)^{3/2} \right). 
\end{align}

\noindent The radiation pressure depends on a cutoff frequency, and we adopt the one used in \cite{cutoff2, cutoff3, cutoff4, SCSconstraints}, $\omega_c \sim \mu^{3/2} I^{-3} (\beta R)^{-1}$. Cosmic string seeded halos have density profiles with $\alpha = 9/4$, therefore we find that the ratio of the gravitational to the radiation pressure force (evaluated at the virial radius) is

\begin{align}
\frac{F_g}{F_r} \approx \, &5 \times 10^8 \left(\frac{R}{t_{eq}}\right)^{5/3} \left(\frac{G\mu }{10^{-10}}\right)^{7/6} \left( \frac{I}{\textrm{GeV}}\right)^{-1} \nonumber \\
&\times \left(\frac{1+z}{1+z_{rec}}\right)^{5/6} \left(1-\left(\frac{1+z}{1+z_{rec}}\right)^{3/2}\right)^{-1}.
\end{align}

\noindent In units of the critical current $I_c = \gamma \kappa^{-1} (G\mu)^{3/2} m_{pl}$, we find

\begin{align}
\frac{F_g}{F_r} \approx \,  &500 \left(\frac{R}{t_{eq}}\right)^{5/3} \left(\frac{G\mu }{10^{-10}}\right)^{-1/3} \left( \frac{I}{I_c}\right)^{-1} \nonumber \\
&\times \left(\frac{1+z}{1+z_{rec}}\right)^{5/6} \left(1-\left(\frac{1+z}{1+z_{rec}}\right)^{3/2}\right)^{-1}.
\end{align}

\noindent The range of parameters we consider above are roughly consistent with the implied values here. For $F_g/F_r > 1$ the gravitational force dominates, and so only in extreme cases of very high currents or very small loops does the additional radiation pressure cause an appreciable effect to the accretion dynamics, thus we neglect it in our analysis. We do, however, use this expression to estimate upper bound on string loop currents in equation (\ref{SCSRP}) of the main text.


\begin{thebibliography}{99}
\bibitem{stellarForm}
Some reference that talks about forming black holes through collapse of a star
\end{thebibliography}
\end{document}